\documentclass[intlimits,twoside,a4paper]{article}
\usepackage{amsmath,amssymb}
\usepackage{graphicx}
\usepackage{color}
\usepackage[T2A]{fontenc}
\usepackage[cp1251]{inputenc}

\usepackage{cmpj2}

\issue{2017}{20}{1}{13201}
\doinumber{10.5488/CMP.20.13201}

\title{Diffusion equations in  inhomogeneous solid having arbitrary gradient concentration}
\author[Y. Bilotsky, M. Gasik, B. Lev]{Y. Bilotsky\refaddr{adr1}, M. Gasik\refaddr{adr1}, B. Lev\refaddr{adr2}}

\addresses{\addr{adr1} Aalto  University, School of Chemical Technology, Materials Processing, PO Box 16200 FI-00076 AALTO Finland
\addr{adr2}  Bogolyubov Institute for Theoretical Physics of the National Academy of Sciences of Ukraine, 14-b~Metrolohichna St., 03143 Kyiv, Ukraine}

\authorcopyright{Y. Bilotsky, M. Gasik, B. Lev, 2017}
\date{Received November 17, 2016, in final form January 6, 2017}

\DeclareMathOperator{\diver}{div}

\begin{document}

\maketitle

\begin{abstract}
A quantum kinetic equation is obtained for an inhomogeneous solid having arbitrary gradient concentration and chemical potential.
We find, starting from nonequilibrium statistical operator, a new equation to describe atom migration in solid states.
In continuous approximation, this equation turns into a non-linear diffusion equation. We derive conditions for which this equation can be reduced to Fick's or Cahn equation.

\pacs 23.23.+x, 56.65.Dy
\end{abstract}

Diffusion phenomena in solid matter have attracted substantial attention for a long time (see as examples \cite{Flyn1,Wei,Flyn2,Fly3,Wer}).
The basis of most approaches to the investigation of these phenomena is expansion in powers of concentration gradients. In our
approach, the atom migrates through grain cross-boundary in solid alloys. The application of this concept to the investigation
of atomic migration trough cross-boundary grain in solid alloys turns out to be unknown so far. Despite the fact that atomic flux through grain
boundary can be moderate, the concentration gradient can be extremely large. Therefore, the linear diffusion theory is
not applicable. This fact has been used in constructing our model. An important assumption in our model is that the concentration of
migrating atoms and the temperature can be described by a quasi-equilibrium distribution function.

This article is based on the well-known fact that diffusion jumping time is longer than other characteristic time scales in a crystal.
The jumping times are normally longer than $10^{-10}$~s even at the melting point, and may become larger at low temperature. Thus, the mean interval
between jumps is longer compared with the phonon oscillation period $\omega^{-1}\sim 10^{-13}$~s as well as with the lifetime of a phonon.
Consequently, dynamic correlations between jumps may be neglected. Therefore, let us start with the Hamiltonian of a system of particles
which interact only with a phonon:
\begin{equation}
\hat{H}=\sum_{i,k}E^{i}_{k}\hat{a}^{+}_{i,k}\hat{a}_{i,k}+\sum_{i,q}\hbar \omega^{i}_{q}\hat{b}^{+}_{i,q}\hat{b}_{i,q}+\hat{H}^{i,j}_{\text{int}}\,,
\end{equation}
where $E^{i}_{k}$ is the energy of a particle and $\hbar \omega^{i}_{q}$ is the energy of a phonon on  $i$ sites, $\hat{a}^{+}_{i,k}$ and $\hat{b}^{+}_{i,q}$
are operators of the birth of a particle and a phonon. The spectrum of energy in different spatial points is different. The Hamiltonian of interacting atoms
and phonons can be written in the form:
\begin{equation}
\hat{H}^{i,j}_{\text{int}}=\sum_{k,q,k',q'}\left(\Phi^{i,j}_{k,q,k',q'}\hat{a}^{+}_{i,k}\hat{b}^{+}_{i,q}\hat{a}_{j,k'}\hat{b}_{j,q'} +\Phi^{+\,i,j}_{k,q,k',q'}\hat{a}^{+}_{i,k}\hat{b}^{+}_{i,q}\hat{a}_{j,k'}\hat{b}_{j,q'}\right),
\end{equation}
where $\Phi^{i,j}_{k,q,k',q'}$ is a matrix element. Interaction between two particles in different locations can be understood as a process for
which one particle interacts with a phonon, receiving additional energy and can jump to another place in a crystal if this place is not occupied
by another particle. If this place is free, a particle can irradiate the other phonon and sit on the free site. Particle number operator
$\hat{N}^{i}=\sum_{k}\hat{a}^{+}_{i,k}\hat{a}_{i,k}$ satisfies the dynamic equation:
\begin{equation}
\dot{\hat{N}}^{i}=-\frac{1}{\ri\hbar}\left[\hat{N}^{i},\hat{H}\right]=\sum_{k,q,k',q'}\left(\Phi^{i,j}_{k,q,k',q'}\hat{a}^{+}_{i,k}\hat{b}^{+}_{i,q}\hat{a}_{j,k'}\hat{b}_{j,q'} -\Phi^{+\,i,j}_{k,q,k',q'}\hat{a}^{+}_{i,k}\hat{b}^{+}_{i,q}\hat{a}_{j,k'}\hat{b}_{j,q'}\right).
\end{equation}
The next step involves the use of Zubarev's non-equilibrium statistical operator
approach \cite{Zub}:
\begin{equation}
\rho=Q^{-1}\exp\left(-\sum_{i}\beta_{i}\hat{H}+\sum_{i}\beta_{i}\mu_{i}\hat{N}^{i}+\sum_{i}\int^{0}_{-\infty}\rd t\,\re^{\epsilon t}\mu_{i}\beta_{i}\dot{\hat{N}}^{i}\right).
\end{equation}
In this case, we will ignore the interaction between particles at distant points. The flux is small but the difference
between temperature or chemical potential can be arbitrary. The time derivative of the average number of
the particles located in the lattice node $i$ comes (by using this non-equilibrium statistical operator) as follows:
\begin{equation}
\langle\dot{\hat{N}}^{i}\rangle= \int^{0}_{-\infty}\rd t\,\re^{\epsilon t}\int^{1}_{0}\big\langle \dot{\hat{N}}^{i}\re^{-\tau \hat{B}}\dot{\hat{N}}^{i}\re^{\tau \hat{B}}\big\rangle_{l}\,,
\label{5}
\end{equation}
where $\langle\ldots\rangle_{l}$ means the average value on the quasi-equilibrium distribution.
\begin{equation}
\hat{\rho}_{l}=Q^{-1}\re^{-\hat{B}}, \qquad \hat{B}=\sum_{i}\beta_{i}\left(\hat{H}^{i}-\mu_{i}\hat{N}^{i}\right).
\end{equation}
The equation~(\ref{5}) is simplified to
\begin{equation}
\langle\dot{\hat{N}}^{i}\rangle= \beta A L_{\dot{\hat{N}}^{i}\dot{\hat{N}}^{i}}
\end{equation}
if the equilibrium distribution is used in the right-hand side of equation~(\ref{5}). Here, $A$ is the difference of chemical potentials in the external and initial states and
\begin{equation}
L_{\dot{N}^{i}\dot{N}^{i}}\equiv \beta^{-1}\int^{0}_{-\infty} \int^{\beta}_{0} \re^{\epsilon t} \big\langle \dot{N}^{i}\dot{N}^{i}(t+\ri\,
\hbar \tau)\big\rangle_{0} \rd t \rd\tau.
\end{equation}
This relation for chemical reactions was obtained in \cite{Yam,Aro}. The equation is valid for small flux and linear dependence
on the thermodynamic force. From the equations
\begin{equation}
\re^{-\tau \hat{B}}\hat{a}_{i,k}\re^{\tau \hat{B}}=\re^{\beta_{i}\left(E^{i}_{k}-\mu_{i}\right)}\hat{a}_{i,k}\,, \qquad \re^{-\tau \hat{B}}\hat{a}^{+}_{i,k}\re^{\tau \hat{B}}=\re^{-\beta_{i}\left(E^{i}_{k}-\mu_{i}\right)}\hat{a}^{+}_{i,k}\,,
\end{equation}
we have
\begin{equation}
\re^{-\tau \hat{B}}\hat{a}^{+}_{i,k}\hat{b}^{+}_{i,q}\hat{a}_{j,k'}\hat{b}_{j,q'}\re^{\tau \hat{B}}=\re^{\left(-\beta_{i}E^{i}_{k}+\beta_{i}E^{j}_{k'}+\beta_{i}\hbar \omega^{i}_{q}-\beta_{i}\hbar \omega^{j}_{q}+\sum_{i}\beta_{i}\mu_{i}\right)\tau}\hat{a}^{+}_{i,k}\hat{b}^{+}_{i,q}\hat{a}_{j,k'}\hat{b}_{j,q'}\,.
\end{equation}
Using this non-equilibrium statistical operator and Wick's theorem one can obtain a kinetic equation for~$N_{i}$~\cite{Zub}
\begin{align}
\dot{N}^{i}&=\sum_{k,q,k',q'}W^{i,i+1}_{k,q,k',q'}\left[-N^{i}_{k}f^{i}_{q}\left(1- N^{i+1}_{k'}\right)\left(1+f^{i+1}_{q'}\right)+N^{i+1}_{k'}f^{i+1}_{q}\left(1- N^{i}_{k}\right)\left(1+f^{i}_{q'}\right)\right]\nonumber\\
&\quad+\sum_{k,q,k',q'}W^{i-1,i}_{k,q,k',q'}\left[-N^{i}_{k'}f^{i}_{q}\left(1- N^{i-1}_{k}\right)\left(1+f^{i-1}_{q'}\right)+N^{i-1}_{k}f^{i-1}_{q}\left(1- N^{i}_{k'}\right)\left(1+f^{i}_{q'}\right)\right],
\end{align}
where
\begin{equation}
W^{i,i+1}_{k,q,k',q'}=\frac{2\pi}{\hbar}\big|\Phi^{i,j}_{k,q,k',q'}\big|^{2}\delta\left(E^{i}_{k}+\hbar \omega^{i}_{q}-E^{j}_{k'}-\hbar \omega^{j}_{q'}\right)
\end{equation}
is the matrix element which takes into account the interaction between two sites, while $f^{i}_{q}$ is distribution function of the
phonon and $N^{i}_{k}$ is distribution function for particles in $i$ site. It is significant that the function $N^{i}_{k}$ on the right-hand side of the equation depends
on two indexes: index $i$ indicates the atom's position in the lattice and index $k$ belongs to the internal degrees of freedom of the atom,
while the derivative of $N^{i}$ on the left-hand side of the equation describes the change on the average numbers of atoms located in the lattice
node $i$ regardless of the index $k$. Let us introduce the coefficient $D_{k}(x)$ as $a^{2}(A^{i}_{k}+A^{i-1}_{k})/2$ and present
$\partial D_{k}(x)/\partial x$ as $a(A^{i}_{k}-A^{i-1}_{k})/a$, where $A^{i+1}_{k}=\sum_{q,k',q'}W^{i,i+1}_{k,q,k',q'}f^{i}_{q}f^{i+1}_{q'}$
in the linear approximation of $n^{i}_{k}$ in a continuum approximation when the lattice constant $a \rightarrow 0$. Only the states of the particle
 involved in diffusion process should be counted in this summation $D_{k}(x)=\frac{1}{2}\sum_{q,q'}\left[W_{k,q,q'}f_{q'}(x+a)+W_{k,q,q'}f_{q'}(x-a)\right]f_{q}(x)$.
For high temperature we can take $f_{q}(x)\approx \frac{1}{\beta(x)\hbar \omega_{q}(x) }$. In the continuous approach, the kinetic equation can be written as:
\begin{equation}
\dot{N}(x)=\frac{\partial}{\partial x}\sum_{k}\hspace{-0.1mm}^{'}D_{k}(x)\frac{\partial N^{k}}{\partial x}\,.
\end{equation}
Therefore, the flux in this case is $J_{N}= \sum_{k}D_{k}(x)\frac{\partial N^{k}}{\partial x}$. If the diffusion coefficient is constant
$D_{k}(x)=\frac{1}{2}\sum_{q,q'}\left[W_{q,q'}f_{q'}(x+a)+W_{q,q'}f_{q'}(x-a)\right]f_{q}(x)\equiv D$, the diffusion equation is reduced
to a well-known simple form:
\begin{equation}
\dot{N}(x)=D \frac{\partial^{2}N}{\partial x^{2}}\,.
\end{equation}
The spatial dependence of the particles number $N_{k}(x)$ can be expressed in terms of the chemical potential and the energy activation as functions of
coordinates. The $ \sum'_{k}N(E_{k})$ takes into account only the states of a particle from which this particle can make a jump in the nearest sites.
The distribution function for high temperature of the particles is of the form $N(E_{k})=\re^{-\beta(x)[E_{k}(x)-\mu(x)]}$. Let us introduce a new
variable $ m_{k}(x)\equiv \re^{-\beta(x)E_{k}(x)}$. Using this variable we can write $N(E_{k})=\re^{\beta(x)\mu(x)}m_{k}(x)$ and the diffusion equation
for the 3D case becomes as follows:
\begin{equation}
\dot{N}(\mathbf{r})=\pmb\nabla \left\{\pmb\nabla \left[\re^{\beta(\mathbf{r})\mu(\mathbf{r})}\right]\sum_{k}D_{k}(\mathbf{r})m_{k}(\mathbf{r})+ \re^{\beta(\mathbf{r})\mu(\mathbf{r})}\sum_{k}D_{k}(\mathbf{r})\pmb\nabla m_{k}(\mathbf{r})\right\}.
\end{equation}
This is an equation of diffusion of a particle in solid matters with different energy, chemical potential and temperature reliefs (the internal degrees of freedom
are taken into account in it). By introducing new functions $D(\mathbf{r})\equiv \sum_{k}D_{k}(\mathbf{r})m_{k}(\mathbf{r})$ and
$\mathbf{V}(\mathbf{r})\equiv \sum_{k}D_{k}(\mathbf{r})\pmb\nabla m_{k}(\mathbf{r})$,  the diffusion equation can be written as follows:
\begin{equation}
\dot{N}(\mathbf{r})=\pmb\nabla \left\{D(\mathbf{r})\re^{\beta(\mathbf{r})\mu(\mathbf{r})}\pmb\nabla [\beta(\mathbf{r})\mu(\mathbf{r})]+\mathbf{V}(\mathbf{r})\re^{\beta(\mathbf{r})\mu(\mathbf{r})}\right\}.
\label{16}
\end{equation}
This equation is correlated with the equation of grain boundary diffusion in a standard form which was obtained in the \cite{Smo}, and the model for diffusion
in a pipe filled with a Knudsen gas  \cite{Kam}. In our case, the external field can be considered as the effect of the inhomogeneous structure of solid matter. This information is included in the coefficient $\mathbf{V}(\mathbf{r}) $ which is dependent on the particle spectrum.

Let us consider some special cases:
\begin{enumerate}
\item Fick's diffusion equation comes from equation~(\ref{16}); if we take into consideration that the diffusion coefficient $D(r)$  varies in space, the chemical potential is the function of concentration only
and $\beta(\mathbf{r})\mu(\mathbf{r})\ll 1$
\begin{equation}
\dot{N}(\mathbf{r})= \pmb\nabla \left\{D\, \pmb\nabla\left[ \re^{\beta(\mathbf{r})\mu(\mathbf{r})}\right]\right\}= \pmb\nabla \left\{D\, \pmb\nabla[ \beta(\mathbf{r})\mu(\mathbf{r})]\right\}.
\end{equation}
If we consider the migration of atoms as a gas with chemical potential  $\mu = \ln \frac{N}{V}\left(\frac{2\pi \hbar^{2}\beta}{m}\right)^{\frac{3}{2}}$,
then this equation is converted into the well-known diffusion equation. This is valid for a small concentration of additional atoms in a solid.

\item Let us consider that the chemical potential is characteristic of the system and is determined by the free energy of the system and by the relation
$\mu =\frac{\delta F}{\delta N}\equiv \frac{1}{V} \frac{\delta F}{\delta c}$, which is the function of concentration $c(\mathbf{r})\equiv \frac{N(\mathbf{r})}{V}$.
In this general approach, we may write the kinetic diffusion equation in the form:
\begin{equation}
\dot{N}(\mathbf{r})=\pmb\nabla \left\{D(\mathbf{r})\re^{\beta(\mathbf{r}) \frac{\delta F(\mathbf{r})}{\delta N}}\pmb\nabla \left[\beta(\mathbf{r}) \frac{\delta F(\mathbf{r})}{\delta N}\right]+\mathbf{V}(\mathbf{r})\re^{\beta(\mathbf{r}) \frac{\delta F(\mathbf{r})}{\delta N}}\right\},
\end{equation}
 where all coefficients vary in space. This equation simplifies greatly if the coefficients $ D $ and $\beta$ are constants.
 This is the well-know diffusion equation \cite{Lan,Gor,Jap}:
\begin{equation}
\dot{N}(\mathbf{r})=D \beta \nabla^{2}\frac{\delta F}{\delta N}\equiv M \nabla^{2}\frac{\delta F}{\delta N}, \qquad M \equiv D \beta.
\end{equation}

\item  The free energy functional for the inhomogeneous system is \cite{Khachaturjan}:
\begin{equation}
F=\int_{V}\left[f(c)+\kappa(\pmb\nabla c)^{2}\right]\rd V,
\end{equation}
which takes into account the arbitrary inhomogeneous distribution of particles. The chemical potential in this case can be determined from the
thermodynamic relation:
\begin{equation}
\mu \equiv \frac{1}{V} \frac{\delta F}{\delta c}= \frac{1}{V} \int_{V}\left[\frac{\partial f(c)}{\partial c}+\frac{\partial \kappa}{\delta c}(\pmb\nabla c)^{2}-2\kappa \nabla^{2}c \right]\rd V.
\end{equation}
This is a famous Cahn nonlinear diffusion equation \cite{Cahn}:
\begin{equation}
\dot{c}(\mathbf{r})= \pmb\nabla M \,\pmb\nabla \mu(\mathbf{r})=M \left[\frac{\partial^{2} f(c)}{\partial c^{2}}\nabla^{2}-2\kappa \nabla^{4}c \right].
\end{equation}

\item  Now, consider a stationary condition
\begin{equation}
\pmb\nabla \left\{D(\mathbf{r})\pmb\nabla \left[\re^{\beta(\mathbf{r}) \frac{\delta F(\mathbf{r})}{\delta N}}\right]+\mathbf{V}(\mathbf{r})\re^{\beta(\mathbf{r}) \frac{\delta F(\mathbf{r})}{\delta N}}\right\}=0,
\end{equation}
therefore, the flux $J=D(\mathbf{r})\pmb\nabla \big[\re^{\beta(\mathbf{r}) \frac{\delta F(\mathbf{r})}{\delta N}}\big]+\mathbf{V}(\mathbf{r})\re^{\beta(\mathbf{r}) \frac{\delta F(\mathbf{r})}{\delta N}}$
is constant and in equilibrium transforms to zero, and thus
\begin{equation}
\pmb\nabla\left[\beta(\mathbf{r})\mu(\mathbf{r})\right]=-\frac{\mathbf{V}(\mathbf{r})}{D(\mathbf{r})}\,.
\end{equation}
This equation has a simple solution
\begin{equation}
\mu(\mathbf{r})=-\frac{1}{\beta(\mathbf{r})}\int^{r}_{0} \frac{\mathbf{V}_{l}(\mathbf{r'})}{D(\mathbf{r'})}\rd x'_{l}\,.
\end{equation}
Here, the chemical potential is a function of concentration.
This approximation can be used to write the Cahn equation for nonzero $\mathbf{V}(\mathbf{r})$
\begin{equation}
M \left[\frac{\partial^{2} f(c)}{\partial c^{2}}\nabla^{2}c-2\kappa \nabla^{4}c \right]+\diver \mathbf{V}=0.
\end{equation}
In the non-equilibrium case, when there exists a stationary flux:
\begin{equation}
D(\mathbf{r})\pmb\nabla \left[\re^{\beta(\mathbf{r}) \frac{\delta F(\mathbf{r})}{\delta N}}\right]+\mathbf{V}(\mathbf{r})\re^{\beta(\mathbf{r}) \frac{\delta F(\mathbf{r})}{\delta N}}=J_{0}(\mathbf{r}),
\end{equation}
there is a solution
\begin{equation}
\exp[\beta(\mathbf{r}) \mu(\mathbf{r})]=S\exp\Bigg[-\int^{r}_{0} \frac{\mathbf{V}_{l}(\mathbf{r'})}{D(\mathbf{r'})}\rd x'_{l}\Bigg]+\exp\Bigg[-\int^{r}_{0} \frac{\mathbf{V}_{l}(\mathbf{r'})}{D(\mathbf{r'})}\rd x'_{l}\Bigg]\cdot\int^{\mathbf{r}}_{0}\frac{J_{0}(\mathbf{r})}{D(\mathbf{r})}\exp\Bigg[+\int^{r}_{0} \frac{\mathbf{V}_{l}(\mathbf{r'})}{D(\mathbf{r'})}\rd x'_{l}\Bigg],
\end{equation}
where $S$ is an arbitrary constant. Thereby, we found a new relationship between temperature, chemical potential, the flux of particles and diffusion coefficient.
\end{enumerate}

In this article we received a new self-consistent equation for a description of the migration of atoms in solid matter by starting
with the quantum non-equilibrium statistical operator \cite{Zub}. This nonlinear differential equation is written in terms of
macroscopic variables, i.e., concentration of migrating atoms. The kinetic characteristics of the system with different (small or
large) gradients of concentrations and inhomogeneous distribution of particles can be calculated using this approach. This
equation is reduced to a diffusion equation or even to Cahn's nonlinear diffusion equation under certain conditions (see above).
Finally, we have to  note that there are several ways to receive macroscopic transport equation starting from quantum
mechanical statistics. See, for example \cite{book1,book2,book3}. We prefer to use Zubaryev's techniques which give a
straightforward result in our case. However, the choice of the method is a matter of one's own opinion.

\ukrainianpart

\title[Рівняння дифузії у твердому тілі]%
{Рівняння дифузії у неоднорідному твердому тілі при довільному градієнті концентрації}

\author{Є. Білоцький\refaddr{adr1},
        М. Гасік\refaddr{adr1},
        Б. Лев\refaddr{adr2}}
\addresses{
\addr{adr1} Університет Аальто, Школа хімічної технології та обробки матеріалів,\\ абонентська скринька 16200, Аальто, Фінляндія
\addr{adr2} Інститут теоретичної фізики ім. М.М. Боголюбова НАН України,\\ вул. Метрологічна, 14-б, 03143 Київ, Україна
}

\makeukrtitle

\begin{abstract}
\tolerance=3000%
Виведено кінетичне рівняння  дифузії і знайдено хімічний потенціал для дифундуючих атомів в неоднорідному твердому тілі при довільному градієнті концентрації.
Базуючись на нерівноважному статистичному операторі, знайдено нове рівняння, що описує міграцію атомів в твердому тілі.
Отримане рівняння в неперервному випадку перетворюється в нелінійне рівняння дифузії, яке при відповідних умовах може бути трансформоване у добре відомі рівняння Фіка або Кана.
Отримано відповідні розв'язки для таких рівнянь.

\end{abstract}
\end{document}